# On mirror symmetry, CSB and anti-hydrogen states in natural atom H.[1]

G. Van Hooydonk, Ghent University, Faculty of Sciences, Krijgslaan 281, B-9000 Ghent (Belgium)
(guido.vanhooydonk@rug.ac.be)

**Abstract**

*Molecular band spectra reveal a left-right symmetry for atoms (Van Hooydonk, Spectrochim. Acta A, 2000, 56, 2273). Intra-atomic left-right symmetry points to anti-atom states and, to make sense, this must also show in line spectra. H Lyman ns½ singlets show a mirror plane at quantum number $n_0 = \frac{1}{2}\pi$. Symmetry breaking oscillator $(1-\frac{1}{2}\pi/n)^2$ means that some of these n-states are anti-hydrogenic. This view runs ahead of CERN's AD-project on antihydrogen.*

A final validation of QED is not yet possible. QED cannot cope with classical chiral symmetry breaking (CSB) effects known from the 19th century. In 2000, we found *that molecular band spectra suggest an atomic left-right symmetry* [1]. If true, this only makes sense if this symmetry is also visible in *atomic line spectra*. So we decided to reanalyze the H Lyman ns½ series with precise data. Kelly's data [2], though useful, have an error of 3 MHz. Erickson [3] claims a better precision with QED calculations, but these lack a classical CSB effect. To test CPT with the CERN AD artificial antihydrogen project [4], data must be as precise as possible. Only H 1s2s is now accurate within parts in $10^{14}$-$10^{15}$, an error of order 10 Hz [5].

Not constant running Rydbergs $R_H(n)$

$$R_H(n) = -E_{nH} \cdot n^2 \quad (1)$$

reflect errors in Bohr theory. The values for Lyman ns½-singlets in Table 1 are plotted versus $1/n$ in Fig. 1. A parabola appears (inconsistent with Dirac theory, due to the Lamb shift). A 2nd order fit

$R_H(n) = 4.36747232754714/n^2 - 5.5556171802571/n$
$\qquad + 109{,}677.585534983 \text{ cm}^{-1}$ (2)

produces a harmonic Rydberg of

$R_{H(harm)} = 109{,}679.3522824 \text{ cm}^{-1} \quad (3)$

different from series limit $E_{1H}$ in Table 1.

Internal anchor (3) is disregarded by NIST, despite its important classical meaning. Strangely enough

$n_0 = 1.572273\ldots \approx \frac{1}{2}\pi \quad (4)$

*is the n-value where the extremum appears. This is close to the generic angle $\frac{1}{2}\pi$ for mirror planes.* For these singlets, a true CSB oscillator, *hidden in QED theory*, appears:

$(1-\frac{1}{2}\pi/n)^2 \quad (5)$

Table 1 – $-E_{nH}$ [3] and $R_H(n)$ for Lyman ns singlets

| n | $-E_{nH}$ [3] | $R_H(n)$ |
|---|---|---|
| 1 | 109678,7737040000 | 109678,773704000 |
| 2 | 27419,8178352300 | 109679,271340920 |
| 3 | 12186,5502372100 | 109678,952134890 |
| 4 | 6854,9188453940 | 109678,701526304 |
| 5 | 4387,1408809200 | 109678,522023000 |
| 6 | 3046,6219504100 | 109678,390214760 |
| 7 | 2238,3324513070 | 109678,290114043 |
| 8 | 1713,7220591550 | 109678,211785920 |
| 9 | 1354,0512214330 | 109678,148936073 |
| 10 | 1096,7809744220 | 109678,097442200 |
| 11 | 906,4302025320 | 109678,054506372 |
| 12 | 761,6529039910 | 109678,018174704 |
| 13 | 648,9821718410 | 109677,987041129 |
| 14 | 559,5814289189 | 109677,960068104 |
| 15 | 487,4574954578 | 109677,936478005 |
| 16 | 428,4293581016 | 109677,915674010 |
| 17 | 379,5082947805 | 109677,897191565 |
| 18 | 338,5119773555 | 109677,880663182 |
| 19 | 303,8168027574 | 109677,865795421 |
| 20 | 274,1946308763 | 109677,852350520 |

*Fit (2) is consistent with QED and with data [2] but (4) and (5) lead to mirror symmetry. The only left-right symmetry imaginable within an atom is atom-antiatom symmetry, and this we already found for atoms in a molecule* [1].

---

Now, (2)-(5) leave a choice between CSB and QED. *The accuracy of closed classical CSB theory must be comparable with that of open complex QED theory* [6]. A harmony obeying $\pi/n$ agrees with the de Broglie equation and since $\alpha/2\pi \approx 2m_e/M_p$, scale factor $\pi$ is like fine structure constant over recoil [6].

In CSB, H is not a 2- but a 4-fermion system [6]. H symmetry is broken with $1\pm m_e/m_H$ ($m_H$ is hydrogen, $m_e$ electron mass). Baryon mass in a reduced mass for a *bound* electron in atom H is 1,836.1526675 $m_e$ (proton mass) for hydrogenic ns-singlets, it is 1,838.1526675 $m_e$ for anti-hydrogenic ns-singlets [6]. The value of $1+2m_e/M_p=1.001089$ is consistent with the observed *free* electron anomaly.

CSB points to different scaling in world and anti-world. Lamb shifts provide the finer details of intra-atomic chiral symmetry breaking.

Fig. 1 $R_H(n)$ versus $1/n$ for H Lyman ns singlets

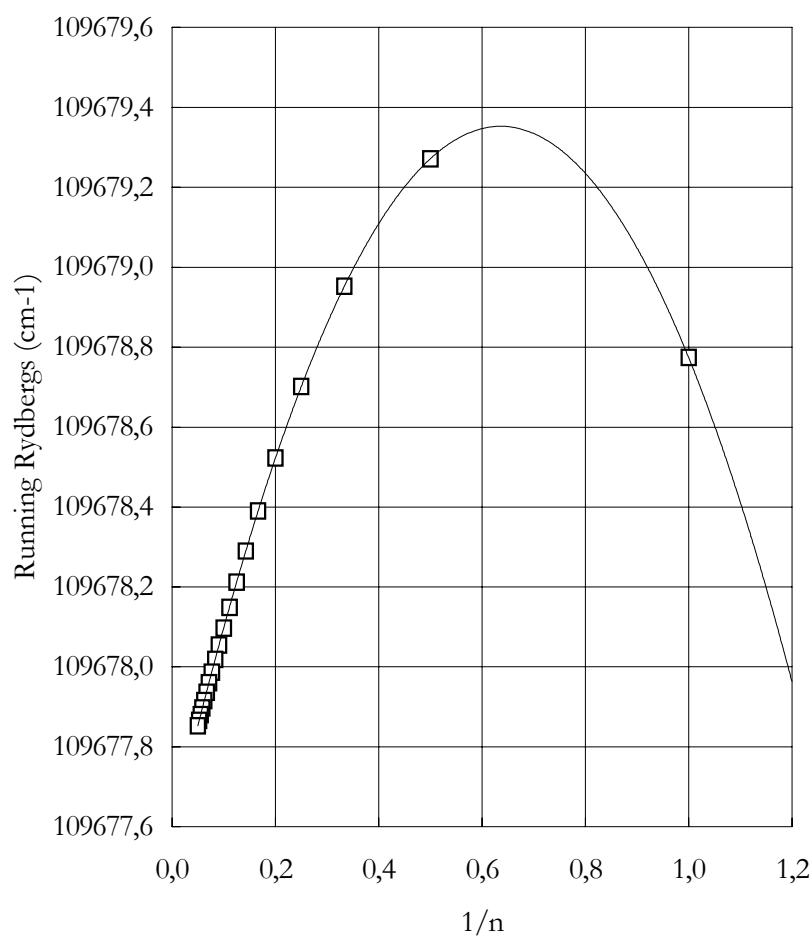